\documentclass[9pt,twocolumn,twoside]{pnas-new-custom}
\templatetype{pnasresearcharticle} 
\usepackage{xcolor}
\usepackage{graphicx} 
\usepackage{threeparttable,booktabs}
\usepackage{makecell}
\usepackage{natbib}

\def\araa{{ARA\string&A}}       
\def\apj{{ApJ}}                 
\def\apjl{{ApJ}}                
\def\apjs{{ApJS}}               
\def\ao{{Appl.~Opt.}}           
\def\aap{{A\string&A}}          
\def\mnras{{MNRAS}}             
\def\nat{{Nature}}              

\newcommand{\affuofa}{Steward Observatory, University of Arizona, 933 North Cherry Avenue, Tucson, AZ 85721, USA}

\begin{document}
\title{JADES: Using NIRCam Photometry to Investigate the Dependence of Stellar Mass Inferences on the IMF in the Early Universe}

\author[a]{Charity Woodrum}
\author[a]{Marcia Rieke}
\author[a]{Zhiyuan Ji}

\author[b,c]{William M.\ Baker}
\author[d]{Rachana Bhatawdekar}
\author[e]{Andrew J.\ Bunker}
\author[f]{St\'ephane Charlot}
\author[g]{Emma Curtis-Lake}
\author[h]{Daniel J.\ Eisenstein}
\author[a]{Kevin Hainline}
\author[i]{Ryan Hausen}
\author[a]{Jakob M.\ Helton}
\author[a,j]{Raphael E.\ Hviding}
\author[h]{Benjamin D.\ Johnson}
\author[k]{Brant Robertson}
\author[a]{Fengwu Sun}
\author[b,c]{Sandro Tacchella}
\author[a]{Lily Whitler}
\author[l]{Christina C.\ Williams}
\author[a]{Christopher N.A.\ Willmer}

\affil[a]{\affuofa}
\affil[b]{Kavli Institute for Cosmology, University of Cambridge, Madingley Road, Cambridge, CB3 0HA, UK} 
\affil[c]{Cavendish Laboratory, University of Cambridge, 19 JJ Thomson Avenue, Cambridge, CB3 0HE, UK} 
\affil[d]{European Space Agency (ESA), European Space Astronomy Centre (ESAC), Camino Bajo del Castillo s/n, 28692 Villanueva de la Cañada, Madrid, Spain} 
\affil[e]{Department of Physics, University of Oxford, Denys Wilkinson Building, Keble Road, Oxford OX1 3RH, UK} 
\affil[f]{Sorbonne Universit\'e, CNRS, UMR 7095, Institut d'Astrophysique de Paris, 98 bis bd Arago, 75014 Paris, France} 
\affil[g]{Centre for Astrophysics Research, Department of Physics, Astronomy and Mathematics, University of Hertfordshire, Hatfield AL10 9AB, UK} 
\affil[h]{Center for Astrophysics $|$ Harvard \& Smithsonian, 60 Garden St., Cambridge, MA 02138, USA} 
\affil[i]{Department of Physics and Astronomy, The Johns Hopkins University, 3400 N. Charles St., Baltimore, MD 21218, USA} 
\affil[j]{Max-Planck-Institut für Astronomie, Königstuhl 17, D-69117, Heidelberg, Germany} 
\affil[k]{Department of Astronomy and Astrophysics, University of California, Santa Cruz, 1156 High Street, Santa Cruz, CA 95064, USA} 
\affil[l]{NSF’s National Optical-Infrared Astronomy Research Laboratory, 950 North Cherry Avenue, Tucson, AZ 85719, USA} 

\leadauthor{Woodrum}

\significancestatement{
The James Webb Space Telescope has enabled the study of the infant universe in unprecedented detail with the hope of revealing how the first galaxies formed and subsequently evolved. If these data were interpreted in the framework of star formation processes in the Milky Way, JWST observations likely contradict cold dark matter theory predictions and would force a reassessment of basic physics.  Here we show that changing star formation parameters to reflect the higher temperatures in the infant universe can resolve the conflict with theory. The cold dark matter paradigm remains consistent with observations.}

\authorcontributions{ CW performed the Prospector analysis, ZJ provided IMF models, AB, FS, and KH provided redshifts, BR, ST, and RH provided photometry and image examination tools, CW, MR, WB, RB, AB, SC, EC-L, DE, JH, REH, BJ, ST, LW, CCW  and CNAW wrote the paper. }
\authordeclaration{The authors have no conflicts to declare.}
\correspondingauthor{\textsuperscript{2} To whom correspondence may be addressed. Email: mrieke@as.arizona.edu}


\begin{abstract}
The detection of numerous and relatively bright galaxies at redshifts $z>9$ has prompted new investigations into the star-forming properties of high-redshift galaxies. Using local forms of the initial mass function (IMF) to estimate stellar masses of these galaxies from their light output leads to galaxy masses that are at the limit allowed for the state of the $\Lambda$CDM Universe at their redshift. We explore how varying the IMF assumed in studies of galaxies in the early universe changes the inferred values for the stellar masses of these galaxies. We infer galaxy properties with the SED fitting code \texttt{Prospector} using varying IMF parameterizations for a sample of 102 galaxies from the JWST Advanced Deep Extragalactic Survey (JADES) spectroscopically confirmed to be at $\mathrm{z>6.7}$, with additional photometry from the JWST Extragalactic Medium Band Survey (JEMS) for twenty-one galaxies. We demonstrate that  models with stellar masses reduced by a factor of three or more do not affect the modeled spectral energy distribution (SED).
\end{abstract}

\dates{Received for review October 6, 2023; compiled \today.}
\doi{\url{www.pnas.org/cgi/doi/10.1073/pnas.XXXXXXXXXX}}

\maketitle
\thispagestyle{firststyle}
\ifthenelse{\boolean{shortarticle}}{\ifthenelse{\boolean{singlecolumn}}{\abscontentformatted}{\abscontent}}{}


\dropcap{T}he first year of observations from the James Webb Space Telescope (JWST) has already revolutionized our understanding of the high redshift Universe. An early paper \citep[ref.][]{Labbe2022} examined the brightest high redshift galaxies and noted that the derived stellar masses are significantly higher (approaching a factor of 10x) than expected at $\sim 600$ million years after the Big Bang. That study used the Salpeter formulation for the initial mass function (IMF), ref. \cite[][]{Salpeter1955}. This result was examined further by ref. \citep{Boylan-Kolchin2023}, using masses from ref. \cite{Labbe2022}, who found that these galaxies lie in a region of the stellar mass density versus stellar mass plot very close to the region where the density of mass in stars exceeds the total baryonic density assuming a $\Lambda$CDM Universe. Ref. \cite{Boylan-Kolchin2023} suggested that the resolution of this tension lies with our lack of knowledge of the star formation process in the early Universe with either star formation being extremely efficient or with a different form for the IMF. Ref. \cite{lovell2023} reached a similar conclusion using a single galaxy and extreme value statistics. Ref. \cite{casey2023}, using data from the COSMOS-Web survey, also find a relatively large number of galaxies with high stellar masses ($\sim 10^{10} M_{\odot}$). These authors assume that an IMF weighted towards high mass stars arises only for PopIII stars. Other evidence suggesting problems with models of early star formation include the discovery of a galaxy with large amounts of carbonaceous dust at an age of $\mathrm{\sim 600}$ million years, ref. \citep{Witstok2023}, a type and quantity of dust that is difficult to produce on this time scale. One possible explanation is dust produced in supernovae from high mass stars rather than the AGB star route which suggests more high mass stars. In the time since ref. \cite{Labbe2022} was published, two of the high mass objects have been observed  spectroscopically, which places them at lower redshift and hence lower mass: Labb\'e ID 13050 moved from z=8.14 to z=5.624, ref. \citep{kocevski2023}, and Labb\'e ID 35300 moved from z=9.08 to z=7.769, ref. \citep{fujimoto2023}. Galaxy 13050 was also shown to be an AGN, which can inflate the inferred mass as the AGN increases the luminosity but not the stellar mass. Removing these two galaxies from  Labb\'e's sample of thirteen galaxies moves the mass density inferred away from the $\Lambda$CDM limit. Note also that neither ref. \cite{robertson_etal23} found such high masses nor did the analysis of the very luminous galaxy GNz11 ref. \citep{Tacchella2023b}. Ref. \cite{bekki2023} found that the high nitrogen abundance observed in GNz11 can be explained using a top-heavy IMF.

Star formation in galaxies at z$\sim$10 must be quite different than local star formation simply because the state of the gas in galaxies at these early times must be different than locally. As discussed below, the gas is warmer and also very likely denser. The star formation rates are likely higher with the study of low redshift galaxies (z$\sim$0.25) suggesting a correlation where galaxies with higher star formation rates have more top-heavy IMFs \citep[ref.][]{gunawardhana2011}. High redshift galaxies may have lower metallicity which may result in a more top-heavy IMF \cite[ref.][]{clauwens2016}. The temperature of the cosmic microwave background is significantly higher at these early times, though this may not be as significant as the increase in dust temperature due to PopII star formation and the presence of silicate dust \cite[ref.][]{derossi2018}. Ref. \cite{vanDokkum2008} presented evidence that for massive elliptical galaxies, even by z$\sim$1, the IMF was already weighted more towards massive stars than the local IMF, but note that ref. \citep{vanDokkum2023} reports on a galaxy at z$\sim 2$ that may have a bottom heavy IMF based on its mass derived from gravitational lensing. 

Ref. \cite{harikane2023} enumerates a number of mechanisms that can modify star formation at high redshifts such as higher gas temperature, which would change the Jeans mass and hence would change the IMF, refs. \citep{vanDokkum2008, Narayanan2012, Jermyn2018, derossi2018}. Higher densities in the gas might mean that disruptive feedback is less effective leading to more efficient star formation. One of the first papers to address the issue of masses and ages of galaxies is ref. \cite{Steinhardt2016}. These authors looked at hierarchical merging in $\Lambda$CDM as well as various possibilities from star formation processes. They emphasized that local star formation processes matched well at least to z$\sim4$ and pointed out that JWST data might be required to resolve what they dubbed as the "Impossibly Early Galaxy Problem." Ref. \cite{Jermyn2018} discusses the importance of the IMF and its relationship to gas temperature as a route to resolving the problem of galaxies apparently being too massive at high redshift. Several recent papers \citep[refs.][]{Sneppen2022,Steinhardt2022,Steinhardt2023} explore how a temperature-dependent IMF results in a top-heavy IMF and how this change to the IMF changes aspects of deriving galaxy properties by template fitting. Their results show that photometric redshifts are essentially unchanged but the inferred masses are reduced by factors of ten or more. 

In this paper we use the deep multi-band photometry from the JWST Advanced Deep Extragalactic Survey (JADES) to explore how much the IMF can be perturbed without affecting the match to the observed SED for a single burst. We find that a top-heavy IMF is consistent with the observed SEDs which is not surprising as the low-mass end of the IMF provides minimal contributions to the luminosity even locally. We use a sample of 102 galaxies from z=6.7 to z=13.2, which have measured spectroscopic redshifts and high quality photometry. We fit the SEDs for these sources using \texttt{Prospector} \cite[v1.1.0, ref.][]{Johnson2021} with modified IMFs and investigate their inferred stellar masses and star formation histories (SFHs). This work is not aimed at deriving the IMF for these galaxies but rather showing that there is a straightforward possible solution to the apparently high masses inferred for similar galaxies. Most likely, a combination of factors controlling the star formation process in the early universe will be required for a complete explanation. See ref. \cite{narayanan2023} for a discussion of how a series of star formation bursts can further complicate stellar mass determination.  We adopt the standard flat $\Lambda$CDM cosmology from Planck18 with $H_0$=67.4 km/s/Mpc and $\Omega_m$=0.315  ref.\citep{Planck2020}.  

\begin{figure}[!htbp]
\includegraphics[width=0.48\textwidth]{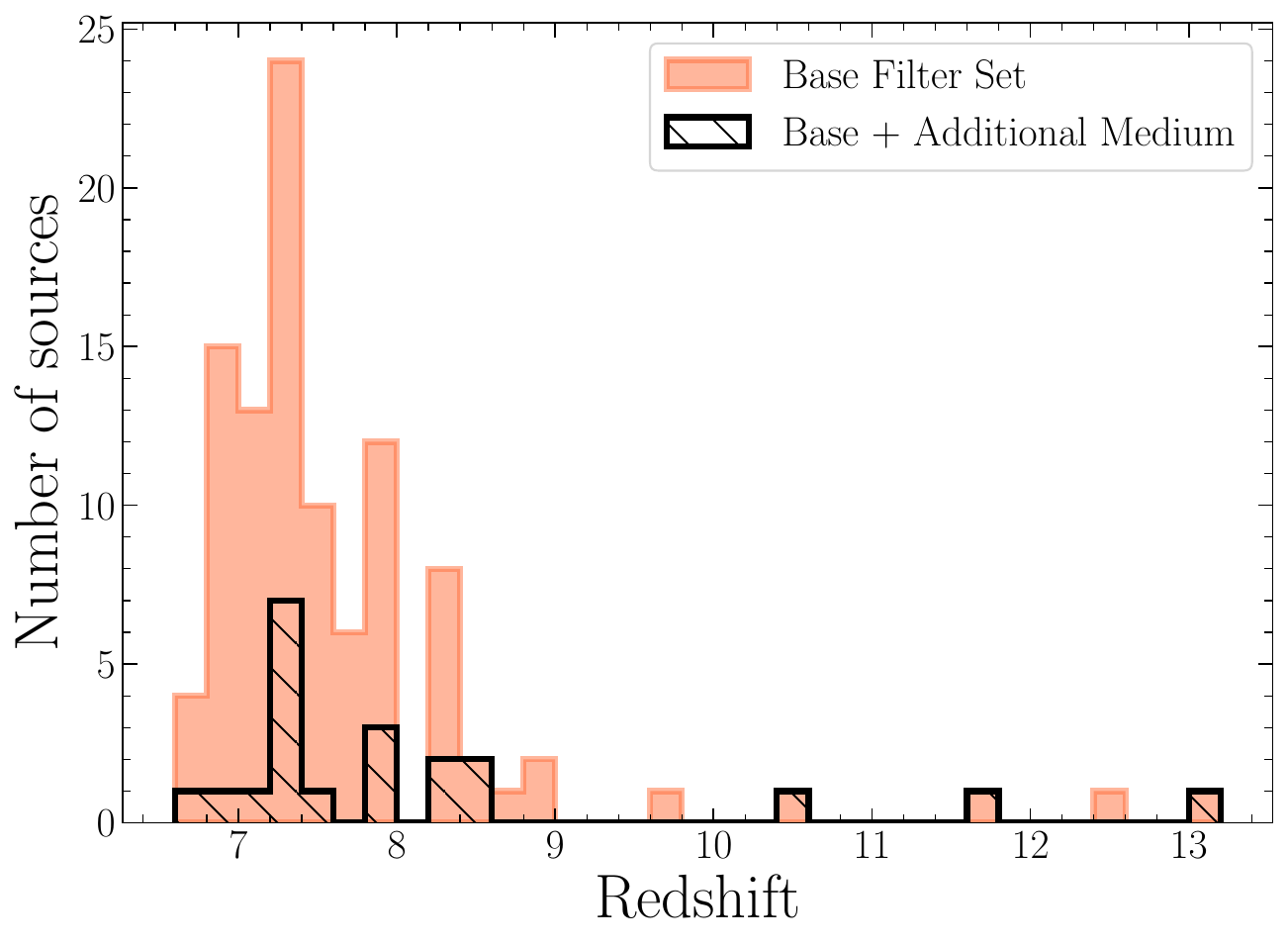}
\caption{The distribution of spectroscopic redshifts for our sample of 102 galaxies from JADES including twenty-one galaxies that have additional medium band photometry from JEMS. \label{fig:redshift}}
\end{figure}

\section*{Data}\label{sec:data}
We utilize deep space-based imaging from JWST with at least eight Near Infrared Camera (NIRCam) photometric bands in the GOODS-S field. The sources all have spectroscopic redshifts measured either as part of the JADES program using the Near Infrared Spectrograph (NIRSpec) refs. \cite{CurtisLake2023, Bunker2023} or as part of the JWST First Reionization Epoch Spectroscopically Complete Observations (FRESCO) (Program ID 1895, PI P. Oesch) program using grisms in NIRCam with redshifts measured by ref. \cite{Sun2023}. The JWST/NIRCam photometry is part of the JWST Advanced Deep Extragalactic Survey ref.\cite{Eisenstein2023, Rieke2023}. The imaging data include F090W, F115W, F150W, F200W, F277W, F356W, F410M, and F444W for the main sample. For a subset of the sample, F182M, F210M, F430M, F460M, and F480M images are also available as part of the JWST Extragalactic Medium-band Survey \citep[JEMS; ref.][]{Williams2023}. As part of the JADES imaging some of these galaxies also have F335M photometry.  The Supplementary Information appendix tabulates the redshifts and photometry for the galaxies used in this study. 

\subsection*{Sample Selection}\label{sec:sample}
The primary sample consists of 102 galaxies from the JADES survey that have confirmed spectroscopic redshifts. Ninety galaxies have FRESCO redshifts, eleven have NIRSpec redshifts, and one galaxy has both. To be included in this sample, the galaxies were also required to have a signal-to-noise ratio greater than five in at least five photometric bands. We select galaxies with $\mathrm{z>6.7}$, the lowest redshift at which [OIII] would be observable in the FRESCO data, which is the main source of our redshifts. This redshift range means that few objects would be detected in F090W because of attenuation in the intergalactic medium \cite[ref.][]{Madau1995}. Any objects with a large ($>0.5$) difference between the spectroscopic redshift and the photometric redshift \cite[ref.][]{Hainline2023} were inspected visually and one object was rejected due to overlapping components. Twenty-one galaxies also have imagery in the medium bands as mentioned above. Results do not differ between the samples with and without JEMS data. The F335M data were acquired as part of the JADES program while the other filters were observed as part of the JEMS program, ref. \cite{Williams2023}. Figure \ref{fig:redshift} shows the distribution of spectroscopic redshifts for our sample. Because of the reliance on FRESCO redshifts, most of the sources in our sample have strong emission lines with the quiescent galaxy found by ref. \cite{Looser2023} as the only object which definitively does not have emission lines. This selection potentially biases our galaxies towards those with vigorous star formation, but currently it is unknown what fraction of $\mathrm{z>6.7}$ galaxies have strong star formation \cite[see also ref.][]{Looser2023b}. Whether the highest redshift portion of our sample, galaxies at $\mathrm{z>9.5}$, have emission lines is indeterminate as $\lambda 5007$\AA [OIII] is redshifted beyond the longest wavelength detected by near-infrared instruments, and detection of shorter wavelength lines which are much weaker requires much higher signal-to-noise than is available now. Ly-$\alpha$ which might be strong is absorbed by the intergalactic medium \cite[e.g., ref.][]{Madau1995}.

\begin{figure}[!htbp]
\includegraphics[width=0.48\textwidth]{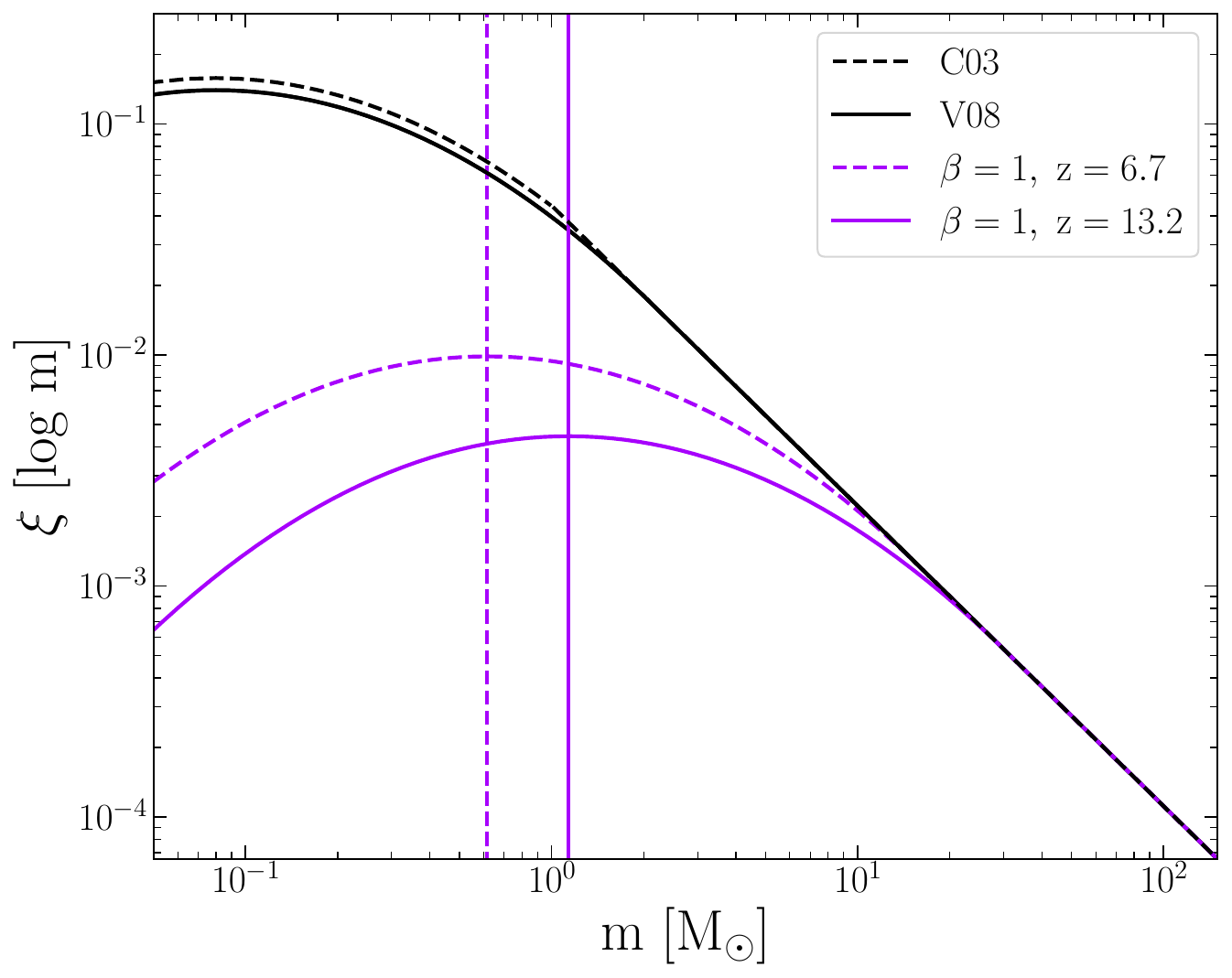} 
\caption{The C03 IMF is shown as a dashed black line compared to the slightly modified form from V08 with $\mathrm{m_c=0.08 M_{\odot}}$ shown as a solid black line. The V08 IMFs with a redshift-dependent characteristic mass, $\mathrm{m_c}\sim(1+z)^{\beta}$, for the minimum and maximum redshifts considered in this paper are shown in purple. The vertical lines indicate the characteristic mass for redshift z=6.7 (dashed purple line) and for redshift z=13.2 (solid purple line). These IMFs use $\mathrm{\beta}$=1.0. \label{fig:imfplot}}
\end{figure}

Photometry was extracted following the procedures outlined in ref. \citep{Rieke2023} with further detail to come from Robertson et al. (in prep.). The \texttt{Prospector} analysis used fluxes derived from images convolved with the PSF for the F444W filter (the PSFs for F460M and F480M are marginally larger but this difference was judged to be insignificant for the analyses here). As described in ref. \citep{Rieke2023}, a Kron radius was determined for each source and fluxes measured for the area defined by the Kron radius. These steps ensure that the same spatial fraction of a galaxy is used across the entire NIRCam wavelength range. 

\section*{Stellar Population Modeling}\label{sec:model}
We fit the photometry with the \texttt{Prospector} \citep[v.1.1.0; ref.][]{Johnson2021} inference framework. \texttt{Prospector} uses the Flexible Stellar Population Synthesis code \citep[\texttt{FSPS; ref.}][]{Conroy2009} via \texttt{python-FSPS} \citep[ref.][]{ForemanMackey2014}. The posterior distributions are sampled using the dynamic nested sampling code \texttt{dynesty}, ref. \citep{dynesty:2020}. 

We use the same physical model as in ref. \cite{Helton2023} following the methodology of ref. \cite{tacchella2022}, with differing IMF prescriptions described in the next section. In brief, the redshifts are fixed at the spectroscopic redshifts shown in Figure \ref{fig:redshift}. We employ the MIST stellar evolutionary tracks and isochrones, refs. \citep{Choi2016, Dotter2016}, which utilizes the MESA stellar evolution package, refs. \citep{Paxton2011, Paxton2013, Paxton2015, Paxton2018}. We use MILES for the stellar spectral library, refs. \citep{Vazdekis2015, Falcon-Barroso2011}. The stellar metallicity, log($Z_*/Z_{\odot}$), was allowed to range from  -2.0 to 0.19. The gas  metallicity, log($Z_{gas}/Z_{\odot}$), was allowed to range from  -2.0 to 0.5.  The IGM absorption is modeled after ref. \cite{Madau1995}, where the overall scaling of the IGM attenuation is a free parameter. For dust attenuation, we assume a flexible attenuation curve with the UV bump tied to the slope of the curve, ref. \citep{Kriek2013}, and a two-component dust model, ref. \citep{Charlot2000}. The nebular emission is based on CLOUDY model grids, ref. \citep{Byler2017}, and includes both nebular continuum and emission line components. The ionization parameter, log(U), was allowed to range from -4 to -1. For the SFH, we use a nonparametric model with the standard continuity prior with six distinct time bins of constant star formation. The bins span from the time of observation to an adopted formation redshift of $\mathrm{z_{form}=20}$. The two most recent age bins are fixed at 0-30 Myr and 30-100 Myr in lookback time in the galaxy's reference frame. The last bin is fixed between 0.85$\mathrm{T_{univ}}$ and $\mathrm{T_{univ}}$, where $\mathrm{T_{univ}}$ is the age of the Universe at the galaxy's spectroscopic redshift, assuming a formation redshift of $\mathrm{z_{form}=20}$. The remaining three bins are spaced evenly in logarithmic time. Changing the SFH prior can also lead to changes in the inferred stellar mass, as has been investigated by refs. \cite{whitler2023}, \cite{tacchella2022}, and \cite{tacchella2023}, however our focus in this paper is on the IMF. Changing the SFH prior mainly affects the amount of mass converted into stars as a function of time with only secondary effects on the total stellar mass for the galaxies at the redshifts in our sample. We note that use of the continuity SFH prior as used in ref. \cite{tacchella2022} does not bias our results as this prior yields stellar masses in the middle of the range for the priors they tested.

\begin{figure*}[!htbp]
\centering
\includegraphics[width=\textwidth]{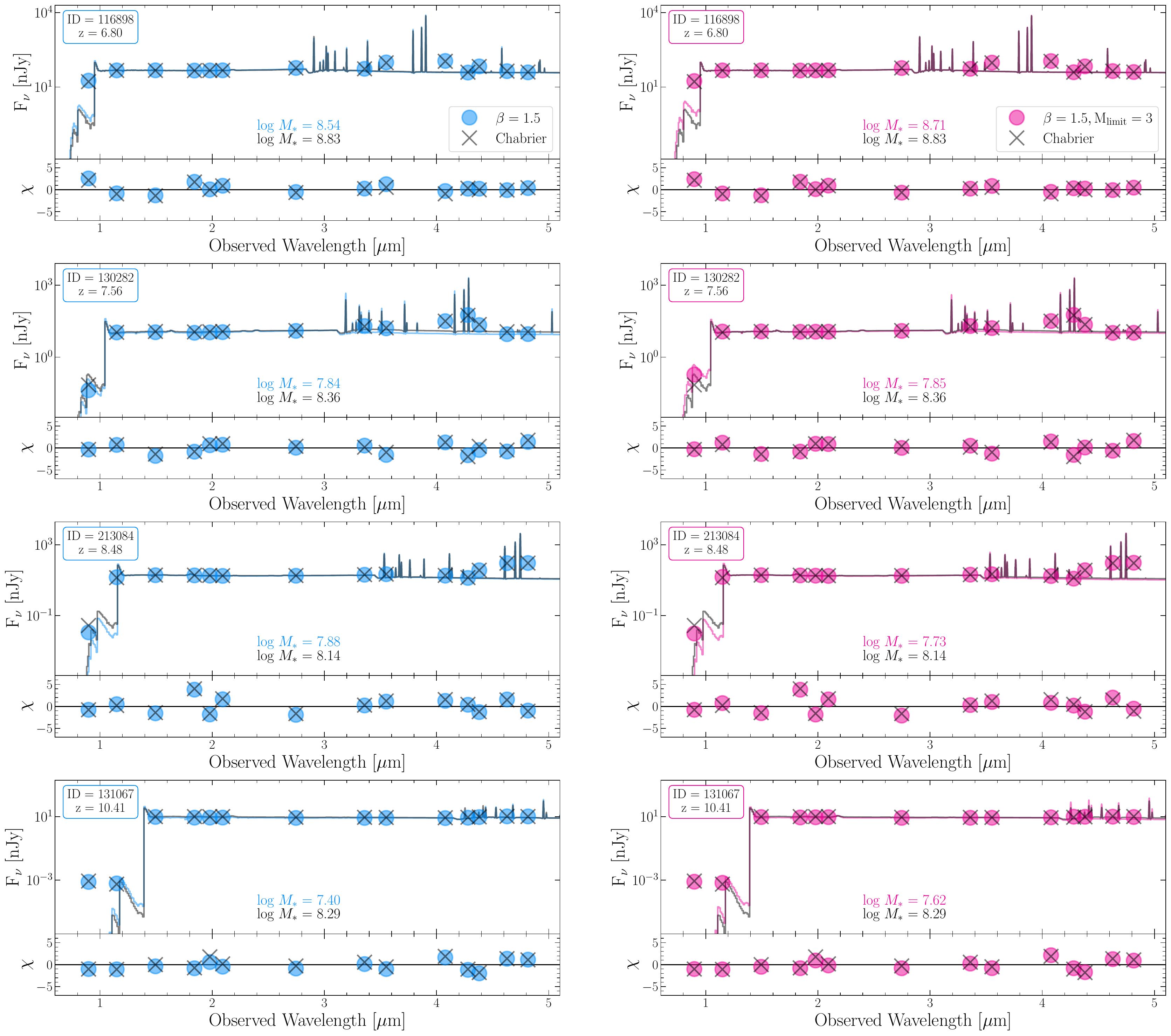}
\caption{A selection of best-fit \texttt{Prospector} SED models for C03 shown in black and for V08 with $\mathrm{\beta=1.5}$ shown in blue in the left column and with $\mathrm{\beta=1.5}$ and $\mathrm{M_{limit}=3}$ in pink in the right column. The smaller panels at the bottom show $\mathrm{\chi}$, defined as $\mathrm{(F_{model}-F_{obs})/\sigma}$, which are nearly identical between models. Therefore, our varying IMF parameterizations produce similar best-fit SED models. \label{fig:sed}}
\end{figure*}

\begin{figure}[!htbp]
\includegraphics[width=0.48\textwidth]{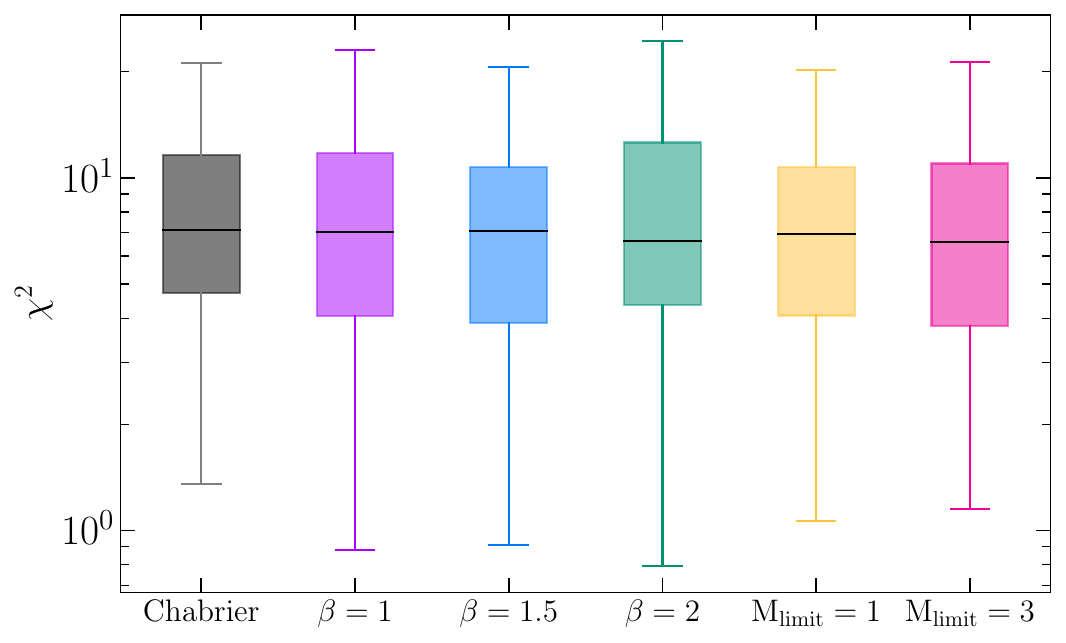}
\caption{Distributions of the $\mathrm{\chi^2}$ statistic for each of the different IMF parameterizations. The box extends from the first quartile to the third quartile with the whiskers showing 1.5 times the inter-quartile range and a line at the median. The data are well-fit by all six of the models. In addition, the $\mathrm{\chi^2}$ statistic is not significantly different between models. Therefore, changing the IMF parameterization does not significantly change the best-fit SED model.\label{fig:chi_squared}}
\end{figure}

\section*{IMF}\label{sec:IMF}
One of the most commonly used parameterizations for the IMF is modeled by a lognormal distribution with a characteristic mass, $\mathrm{m_c}$, \cite[ref.][hereafter C03]{Chabrier2003}. van Dokkum 2008, \cite[ref.][hereafter V08]{vanDokkum2008}, introduced a slightly modified form of the C03 IMF which allows for a varying $\mathrm{m_c}$, where $\mathrm{m_c=0.08\ M_{\odot}}$ is almost identical to the C03 IMF. The functional form of V08 is:
\begin{alignat}{2}
&\xi(m) = 
\begin{cases} 
A_l(0.5 n_c m_c)^{-x} \rm{exp} \left[ - \frac{(\rm{log}\ m - \rm{log}\ m_c)^2}{2\sigma^2} \right], &m\leq n_c m_c, \\
A_h m^{-x}, &m>n_c m_c,
\end{cases}
\end{alignat}
with $A_l=0.140$, $n_c=25$, $\sigma=0.69$, $A_h=0.0443$, and x=1.3 with x refered to as the slope of the IMF. This formulation of the IMF allows the characteristic turnover mass, where the IMF begins to decline, to vary with redshift.

As mentioned in the introduction, the characteristic mass, $\mathrm{m_c}$, may change with the temperature of the ISM. If we assume the ISM temperature of galaxies scales with the temperature of the CMB and purely based on a Jeans argument ref. \citep{Jeans1902}, then $\mathrm{m_c \sim (T_{ISM})^{1.5} \sim (T_{CMB})^{1.5} \sim (1+z)^{1.5}}$. Other studies have suggested different scale factors. For example, ref. \cite{Hopkins2012} showed that $\mathrm{m_c \sim (1+z)}$ and ref. \cite{Steinhardt2020} argue that $\mathrm{m_c \sim (1+z)^2}$. Ref. \cite{sommovigo2022} use ALMA data to show that the dust temperature over the range up to z$\sim$7 increases as $(1+z)^{0.42}$ which supports an increasing value for $\mathrm{m_c}$. The functional form of the $\mathrm{m_c}$ to z relation will be more complicated than just the relation for dust temperature because of other factors, such as gas density, which also play a role in setting the mass of collapsing clouds. We modify the V08 IMF by making $\mathrm{m_c}$ proportional to  $\mathrm{(1+z)^{\beta}}$, where $\mathrm{\beta}$=1, 1.5 and 2 which renders the IMF redshift-dependent. In addition, we use the V08 IMF with $\mathrm{\beta=1.5}$ and apply a lower limit to the IMF mass. The default lower limit is $\mathrm{M_{limit}=0.08 M_{\odot}}$ which we change to $\mathrm{M_{limit}=1 M_{\odot}\ and\ 3 M_{\odot}}$. We note that the default upper limit on the IMF mass is 120$\mathrm{M_{\odot}}$, which we keep unchanged. The effect of changing the upper mass limit would likely result is similar mass estimates but with younger ages.  Figure \ref{fig:imfplot} compares one of our modfied IMFs at two redshifts to the original V08 version of the C03 IMF.

\section*{Results}
In this section, we present the inferred physical properties of galaxies in our sample with differing IMF parameterizations described in the previous section. All values are reported as the median, with uncertainties as the 16th and 84th percentiles of the posterior probability. For the quiescent galaxy in our sample, we check if our fitting results are consistent with those listed in ref. \cite{Looser2023}. We compare inferred values from our model with the C03 IMF parameterization and their model with the same SFH prior used in this work (the standard continuity prior). We find that all of the inferred parameters are consistent with each other within uncertainties. The inferred \texttt{Prospector} properties are included in the Supplementary Information appendix.

In Figure \ref{fig:sed}, we show examples of the best-fit SEDs for the C03 models compared to the V08 models with $\mathrm{\beta=1.5}$ and also with $\mathrm{\beta=1.5}$ and $\mathrm{M_{limit=3}}$. The residuals, defined as $\mathrm{\chi = (F_{model}-F_{obs})/\sigma_{obs}}$, are centered around 0 and show that the data are well-fit by the model. In addition, the residuals among the different models are nearly identical. 

To determine if the data are better fit by one model over the other, we calculate the $\mathrm{\chi^2}$ statistic using the best-fit model photometry as $\mathrm{\chi^2 = \sum (F_{model}-F_{obs})^2/\sigma_{obs}^2}$, where $\mathrm{F_{model}}$ and $\mathrm{F_{obs}}$ are the observed and model fluxes, respectively, and $\mathrm{\sigma_{obs}}$ is the observed photometric uncertainty, see Figure \ref{fig:chi_squared}. We find that the data are well-fit by all six of the models. In addition, the $\mathrm{\chi^2}$ statistic is not significantly different between models. Therefore, changing the IMF parameterization results in model fits
that match the observed SEDs equally well.

\begin{figure*}[!htbp]
\includegraphics[width=\textwidth]{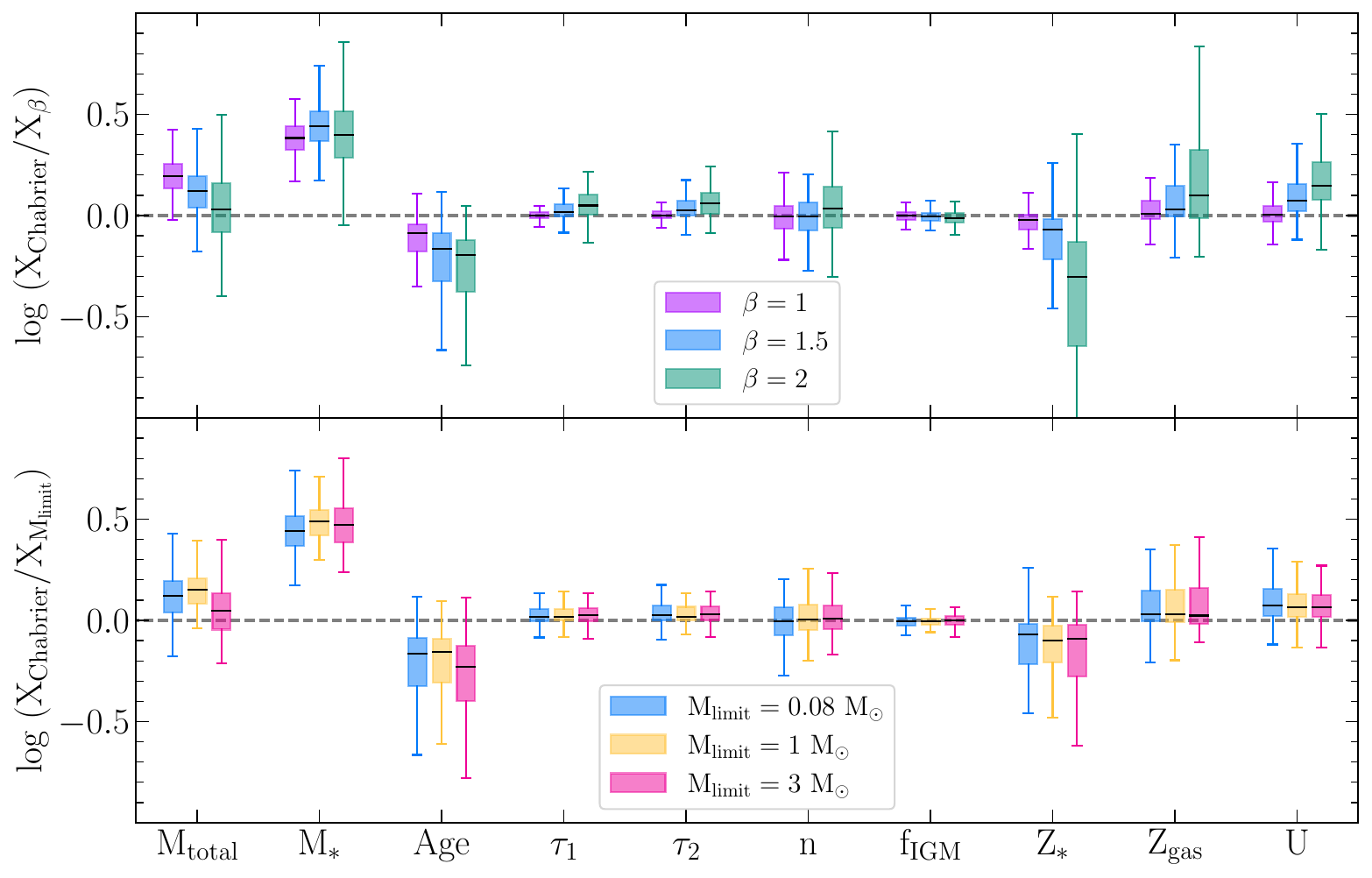}
\caption{Distributions of the differences between the parameters inferred with our varying IMF parameterizations. The box and whisker ranges are the same as in Figure \ref{fig:chi_squared}. The inferred parameters shown here include the total formed mass ($\mathrm{M_{total}}$), the surviving stellar mass ($\mathrm{M_*}$), the mass-weighted age, the birth-cloud dust ($\mathrm{\tau_1}$), the diffuse dust ($\mathrm{\tau_2}$), the power-law modifier to the shape of the dust attenuation curve (n), the factor used to scale the IGM attenuation curve ($\mathrm{f_{IGM}}$), the stellar metallicity ($\mathrm{Z_*}$), the gas-phase metallicity ($\mathrm{Z_{gas}}$), and the ionization parameter for nebular emission (U). The models with varying lower mass limits all use $\beta$ = 1.5. Compared to the values inferred using V08 with varying scale factors for $\mathrm{m_c}$ and varying lower limits on the IMF mass, the inferred C03 values are $\mathrm{\approx 0.1-0.2\ dex}$ higher for the total formed mass ($\mathrm{M_{total}}$), $\mathrm{\approx 0.4-0.5\ dex}$ higher for the surviving stellar mass ($\mathrm{M_*}$), and $\mathrm{\approx 0.1-0.2\ dex}$ lower for the mass-weighted age. \label{fig:param_compare}}
\end{figure*}

\begin{figure*}[!htbp]
\includegraphics[width=\textwidth]{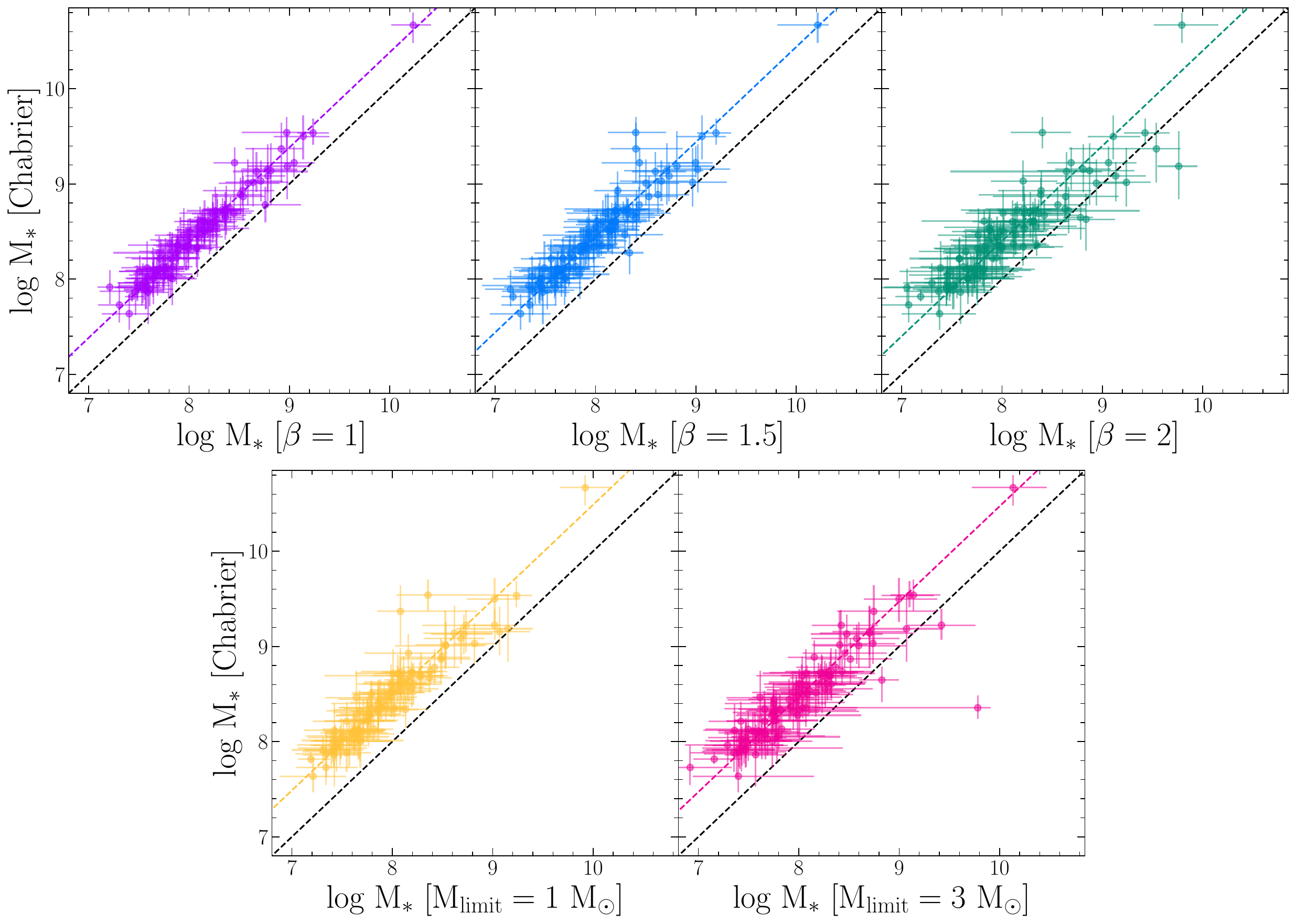}
\caption{Comparison of stellar masses in units of $\mathrm{M_{\odot}}$ determined using a C03 IMF with varying versions of the V08 IMF. For the latter, we let the characteristic mass ($\mathrm{m_c}$) scale by $\mathrm{(1+z)^{\beta}}$. In addition, we place lower limits on the IMF mass ($\mathrm{M_{limit}}$). The median offset is shown as a colored dashed line for the different IMF parameterizations. \label{fig:mass_compare}}
\end{figure*}

Next, we compare the distributions of the differences between the inferred galaxy properties using the C03 model and the varying V08 models, see Figure \ref{fig:param_compare}. The inferred parameters include the total formed mass ($\mathrm{M_{total}}$), the surviving stellar mass ($\mathrm{M_*}$), the mass-weighted age, the birth-cloud dust attenuation ($\mathrm{\tau_1}$), the diffuse dust attenuation ($\mathrm{\tau_2}$), the power-law modifier to the shape of the dust attenuation curve (n), the factor used to scale the IGM attenuation curve ($\mathrm{f_{IGM}}$), the stellar metallicity ($\mathrm{Z_*}$), the gas-phase metallicity ($\mathrm{Z_{gas}}$), and the ionization parameter for nebular emission (U). Compared to the values inferred using V08 with varying scale factors for $\mathrm{m_c}$ and varying lower limits on the IMF masses, the inferred median C03 values are $\mathrm{\approx 0.1-0.2\ dex}$ higher for the total formed mass ($\mathrm{M_{total}}$), $\mathrm{\approx 0.4-0.5\ dex}$ higher for the surviving stellar mass ($\mathrm{M_*}$), and $\mathrm{\approx 0.1-0.2\ dex}$ lower for the mass-weighted age. The most significant differences are between the stellar masses, which we highlight in Figure \ref{fig:mass_compare} and we list the median offsets in Table \ref{tab:offsets}.

\begin{table}[!htbp]
\caption{$\mathrm{M_*}$ Offsets \label{tab:offsets}}
\begin{tabular*}{\hsize}{@{\extracolsep{\fill}}ll}
IMF & Mass Reduction Factor \cr
\hline
$\mathrm{\beta=1}$ & 2.4 \\
$\mathrm{\beta=1.5}$ & 2.8 \\
$\mathrm{\beta=2}$ & 2.5 \\
$\mathrm{M_{limit}=1\ M_{\odot}}$ & 3.1 \\
$\mathrm{M_{limit}=3\ M_{\odot}}$ & 3.0 \\
\hline
\end{tabular*}
\addtabletext{The median offset between stellar masses inferred from the C03 parameterization and the differing V08 parameterizations, shown as dashed lines in Figure \ref{fig:mass_compare}.}
\end{table}

In summary, varying the IMF parameterization results in SED models that are not substantially different from one another. However, their stellar masses can differ significantly, with over three times smaller inferred stellar masses than for the commonly used C03 model. In addition, the mass-weighted ages are lower for the C03 model, meaning that the C03 model infers SFHs that form larger masses over a shorter amount of time. As mentioned in the introduction, a variety of factors can influence stellar mass estimates derived from SED fitting to observed fluxes. This study is confined to examining how much plausible changes to the IMF change the derived masses, and the factor of three reduction found here, Table \ref{tab:offsets}, would significantly reduce the tension with the allowed amount of stellar mass in $\Lambda$CDM models.

\section*{Discussion}
Our results indicate that the changes in the IMF that are likely for high redshift star formation can reduce the stellar mass inferred from galaxy photometry by as much as a factor of three as compared to the mass inferred from use of the local C03 IMF. In ref. \citep{Labbe2022}, masses were inferred using the Salpeter IMF which does not have a low mass turnover and is rarely used now in Milky Way star formation studies. A Salpeter IMF yields a mass yet another factor of  $\sim 2$ higher mass than the C03 IMF which does have a low mass turnover as illustrated in Figure \ref{fig:imfplot}. A total reduction of a factor of 6 in mass by appropriate choice of IMF (see Table \ref{tab:offsets}) for the  Labb\'e sample could be achieved. While nearly all of the galaxies in our sample are less massive and bluer than the galaxies in the Labb\'e sample, one galaxy, JADES-GS-53.16137-27.73766 at z=6.81, would have been selected using their criteria. The IMF-modfied fits for this galaxy reduce the mass inferred from C03 by a factor of 3.4 showing that our suggestion of masses being smaller applies to the double-break galaxies defined in ref. \citep{Labbe2022}.  This IMF-based mass reduction along with the correction of two redshifts in the Labb\'e sample removes the need for the extremely high star formation efficiencies derived by ref. \cite{Boylan-Kolchin2023}. A similar conclusion is reached by ref. \cite{Steinhardt2023}. 

The galaxies used in this study all have spectroscopic redshifts so there are no uncertainties on the light travel time from them. The redshift interval from z$\sim$13.2 to z$\sim6.7$ corresponds to ages of 320 million years and 804 million years after the Big Bang, respectively. In terms of stellar evolution, these ages correspond to changes in main sequence turn-off ranging from O and B stars to F stars depending on when the stars first formed. The picture that is developing for these high redshift objects is one with strong on-going star formation as evidenced by the strong emission lines detected, and which are present in nearly all of the galaxies in our sample. The output of high mass stars capable of ionizing the ISM completely hides the low mass end of the mass function so it is not surprising that our \texttt{Prospector} models are so insensitive to the parameterization of the IMF. What is clear is that the high redshift galaxy population being discovered in JWST data is more luminous than expected, refs. \cite[e.g.,][]{Finkelstein2022a, robertson2022, stark2016}, but not necessarily more massive. Because of the many reasons for the high redshift IMF to differ from C03, this component of minimizing the tension with $\Lambda$CDM needs to be taken into account. However, a complete understanding of the SFHs of the galaxies at early times awaits more detailed spectroscopy. The solution to measuring galaxy masses accurately will require high spectral resolution data that can be used for measurement of dynamical masses although such data will only provide upper limits on the stellar mass.

\subsection*{Supporting Information Appendix (SI)}

The spreadsheet tabulates the redshifts, photometry, and \texttt{Prospector} inferred parameters for the galaxies used in this study.



\acknow{CW, MR, ZJ, JH, FS, CNAW, KH, LW, RH, DE, BJ, and BR are supported by NASA contract NAS5-02105 to the University of Arizona. AJB has received funding from the European Research Council (ERC) under the European Union’s Horizon 2020 Advanced Grant 789056 “First Galaxies.” WB acknowledges support by the Science and Technology Facilities Council (STFC), ERC Advanced Grant 695671 "QUENCH", and by the UKRI Frontier Research grant RISEandFALL. ECL acknowledges support of an STFC Webb Fellowship (ST/W001438/1). REH acknowledges support from the National Science Foundation Graduate Research Fellowship Program under Grant No. DGE-1746060. The research of CCW is supported by NOIRLab, which is managed by the Association of Universities for Research in Astronomy (AURA) under a cooperative agreement with the National Science Foundation. RH acknowledges funding provided by the Johns Hopkins University, Institute for Data Intensive Engineering and Science (IDIES).

This material is based upon High Performance Computing (HPC) resources supported by the University of Arizona TRIF, UITS, and Research, Innovation, and Impact (RII) and maintained by the UArizona Research Technologies department. We respectfully acknowledge the University of Arizona is on the land and territories of Indigenous peoples. Today, Arizona is home to 22 federally recognized tribes, with Tucson being home to the O'odham and the Yaqui. Committed to diversity and inclusion, the University strives to build sustainable relationships with sovereign Native Nations and Indigenous communities through education offerings, partnerships, and community service.}

\showacknow{} 




\begin{thebibliography}{}
\expandafter\ifx\csname natexlab\endcsname\relax\def\natexlab#1{#1}\fi
\providecommand{\url}[1]{\href{#1}{#1}}
\providecommand{\dodoi}[1]{doi:~\href{http://doi.org/#1}{\nolinkurl{#1}}}
\providecommand{\doeprint}[1]{\href{http://ascl.net/#1}{\nolinkurl{http://ascl.net/#1}}}
\providecommand{\doarXiv}[1]{\href{https://arxiv.org/abs/#1}{\nolinkurl{https://arxiv.org/abs/#1}}}

\bibitem[{{Bekki} \& {Tsujimoto}(2023)}]{bekki2023}
{Bekki}, K., \& {Tsujimoto}, T. 2023, \mnras, 526, L26,
  \dodoi{10.1093/mnrasl/slad108}

\bibitem[{{Boylan-Kolchin}(2023)}]{Boylan-Kolchin2023}
{Boylan-Kolchin}, M. 2023, Nature Astronomy, 7, 731,
  \dodoi{10.1038/s41550-023-01937-7}

\bibitem[{{Bunker} {et~al.}(2023){Bunker}, {Cameron}, {Curtis-Lake},
  {Jakobsen}, {Carniani}, {Curti}, {Witstok}, {Maiolino}, {D'Eugenio},
  {Looser}, {Willott}, {Bonaventura}, {Hainline}, {Uebler}, {Willmer},
  {Saxena}, {Smit}, {Alberts}, {Arribas}, {Baker}, {Baum}, {Bhatawdekar},
  {Bowler}, {Boyett}, {Charlot}, {Chen}, {Chevallard}, {Circosta}, {DeCoursey},
  {de Graaff}, {Egami}, {Eisenstein}, {Endsley}, {Ferruit}, {Giardino},
  {Hausen}, {Helton}, {Hviding}, {Ji}, {Johnson}, {Jones}, {Kumari}, {Laseter},
  {Luetzgendorf}, {Maseda}, {Nelson}, {Parlanti}, {Perna}, {Rawle}, {Rix},
  {Rieke}, {Robertson}, {Rodriguez Del Pino}, {Sandles}, {Scholtz}, {Sharpe},
  {Skarbinski}, {Stark}, {Sun}, {Tacchella}, {Topping}, {Villanueva},
  {Wallace}, {Williams}, \& {Woodrum}}]{Bunker2023}
{Bunker}, A.~J., {Cameron}, A.~J., {Curtis-Lake}, E., {et~al.} 2023, arXiv
  e-prints, arXiv:2306.02467, \dodoi{10.48550/arXiv.2306.02467}

\bibitem[{{Byler} {et~al.}(2017){Byler}, {Dalcanton}, {Conroy}, \&
  {Johnson}}]{Byler2017}
{Byler}, N., {Dalcanton}, J.~J., {Conroy}, C., \& {Johnson}, B.~D. 2017, \apj,
  840, 44, \dodoi{10.3847/1538-4357/aa6c66}

\bibitem[{{Casey} {et~al.}(2023){Casey}, {Akins}, {Shuntov}, {Ilbert},
  {Paquereau}, {Franco}, {Hayward}, {Finkelstein}, {Boylan-Kolchin},
  {Robertson}, {Allen}, {Brinch}, {Cooper}, {Ding}, {Drakos}, {Faisst},
  {Fujimoto}, {Gillman}, {Harish}, {Hirschmann}, {Jin}, {Kartaltepe},
  {Koekemoer}, {Kokorev}, {Liu}, {Long}, {Magdis}, {Maraston}, {Martin},
  {McCracken}, {McKinney}, {Mobasher}, {Rhodes}, {Rich}, {Sanders},
  {Silverman}, {Toft}, {Vijayan}, {Weaver}, {Wilkins}, {Yang}, \&
  {Zavala}}]{casey2023}
{Casey}, C.~M., {Akins}, H.~B., {Shuntov}, M., {et~al.} 2023, arXiv e-prints,
  arXiv:2308.10932, \dodoi{10.48550/arXiv.2308.10932}

\bibitem[{{Charlot} \& {Fall}(2000)}]{Charlot2000}
{Charlot}, S., \& {Fall}, S.~M. 2000, \apj, 539, 718, \dodoi{10.1086/309250}

\bibitem[{{Choi} {et~al.}(2016){Choi}, {Dotter}, {Conroy}, {Cantiello},
  {Paxton}, \& {Johnson}}]{Choi2016}
{Choi}, J., {Dotter}, A., {Conroy}, C., {et~al.} 2016, \apj, 823, 102,
  \dodoi{10.3847/0004-637X/823/2/102}

\bibitem[{{Clauwens} {et~al.}(2016){Clauwens}, {Schaye}, \&
  {Franx}}]{clauwens2016}
{Clauwens}, B., {Schaye}, J., \& {Franx}, M. 2016, \mnras, 462, 2832,
  \dodoi{10.1093/mnras/stw1808}

\bibitem[{{Conroy} {et~al.}(2009){Conroy}, {Gunn}, \& {White}}]{Conroy2009}
{Conroy}, C., {Gunn}, J.~E., \& {White}, M. 2009, \apj, 699, 486,
  \dodoi{10.1088/0004-637X/699/1/486}

\bibitem[{{Curtis-Lake} {et~al.}(2023){Curtis-Lake}, {Carniani}, {Cameron},
  {Charlot}, {Jakobsen}, {Maiolino}, {Bunker}, {Witstok}, {Smit}, {Chevallard},
  {Willott}, {Ferruit}, {Arribas}, {Bonaventura}, {Curti}, {D'Eugenio},
  {Franx}, {Giardino}, {Looser}, {L{\"u}tzgendorf}, {Maseda}, {Rawle}, {Rix},
  {Rodr{\'\i}guez del Pino}, {{\"U}bler}, {Sirianni}, {Dressler}, {Egami},
  {Eisenstein}, {Endsley}, {Hainline}, {Hausen}, {Johnson}, {Rieke},
  {Robertson}, {Shivaei}, {Stark}, {Tacchella}, {Williams}, {Willmer},
  {Bhatawdekar}, {Bowler}, {Boyett}, {Chen}, {de Graaff}, {Helton}, {Hviding},
  {Jones}, {Kumari}, {Lyu}, {Nelson}, {Perna}, {Sandles}, {Saxena}, {Suess},
  {Sun}, {Topping}, {Wallace}, \& {Whitler}}]{CurtisLake2023}
{Curtis-Lake}, E., {Carniani}, S., {Cameron}, A., {et~al.} 2023, Nature
  Astronomy, 7, 622, \dodoi{10.1038/s41550-023-01918-w}

\bibitem[{{De Rossi} {et~al.}(2018){De Rossi}, {Rieke}, {Shivaei}, {Bromm}, \&
  {Lyu}}]{derossi2018}
{De Rossi}, M.~E., {Rieke}, G.~H., {Shivaei}, I., {Bromm}, V., \& {Lyu}, J.
  2018, \apj, 869, 4, \dodoi{10.3847/1538-4357/aaebf8}

\bibitem[{{Dotter}(2016)}]{Dotter2016}
{Dotter}, A. 2016, \apjs, 222, 8, \dodoi{10.3847/0067-0049/222/1/8}

\bibitem[{{Eisenstein} {et~al.}(2023){Eisenstein}, {Willott}, {Alberts},
  {Arribas}, {Bonaventura}, {Bunker}, {Cameron}, {Carniani}, {Charlot},
  {Curtis-Lake}, {D'Eugenio}, {Endsley}, {Ferruit}, {Giardino}, {Hainline},
  {Hausen}, {Jakobsen}, {Johnson}, {Maiolino}, {Rieke}, {Rieke}, {Rix},
  {Robertson}, {Stark}, {Tacchella}, {Williams}, {Willmer}, {Baker}, {Baum},
  {Bhatawdekar}, {Boyett}, {Chen}, {Chevallard}, {Circosta}, {Curti},
  {Danhaive}, {DeCoursey}, {de Graaff}, {Dressler}, {Egami}, {Helton},
  {Hviding}, {Ji}, {Jones}, {Kumari}, {L{\"u}tzgendorf}, {Laseter}, {Looser},
  {Lyu}, {Maseda}, {Nelson}, {Parlanti}, {Perna}, {Pusk{\'a}s}, {Rawle},
  {Rodr{\'\i}guez Del Pino}, {Sandles}, {Saxena}, {Scholtz}, {Sharpe},
  {Shivaei}, {Silcock}, {Simmonds}, {Skarbinski}, {Smit}, {Stone}, {Suess},
  {Sun}, {Tang}, {Topping}, {{\"U}bler}, {Villanueva}, {Wallace}, {Whitler},
  {Witstok}, \& {Woodrum}}]{Eisenstein2023}
{Eisenstein}, D.~J., {Willott}, C., {Alberts}, S., {et~al.} 2023, arXiv
  e-prints, arXiv:2306.02465, \dodoi{10.48550/arXiv.2306.02465}

\bibitem[{{Falc{\'o}n-Barroso} {et~al.}(2011){Falc{\'o}n-Barroso},
  {S{\'a}nchez-Bl{\'a}zquez}, {Vazdekis}, {Ricciardelli}, {Cardiel}, {Cenarro},
  {Gorgas}, \& {Peletier}}]{Falcon-Barroso2011}
{Falc{\'o}n-Barroso}, J., {S{\'a}nchez-Bl{\'a}zquez}, P., {Vazdekis}, A.,
  {et~al.} 2011, \aap, 532, A95, \dodoi{10.1051/0004-6361/201116842}

\bibitem[{{Finkelstein} {et~al.}(2022){Finkelstein}, {Bagley}, {Song},
  {Larson}, {Papovich}, {Dickinson}, {Finkelstein}, {Koekemoer}, {Pirzkal},
  {Somerville}, {Yung}, {Behroozi}, {Ferguson}, {Giavalisco}, {Grogin},
  {Hathi}, {Hutchison}, {Jung}, {Kocevski}, {Kawinwanichakij}, {Rojas-Ruiz},
  {Ryan}, {Snyder}, \& {Tacchella}}]{Finkelstein2022a}
{Finkelstein}, S.~L., {Bagley}, M., {Song}, M., {et~al.} 2022, \apj, 928, 52,
  \dodoi{10.3847/1538-4357/ac3aed}

\bibitem[{{Foreman-Mackey} {et~al.}(2014){Foreman-Mackey}, {Sick}, \&
  {Johnson}}]{ForemanMackey2014}
{Foreman-Mackey}, D., {Sick}, J., \& {Johnson}, B. 2014, {python-fsps: Python
  bindings to FSPS (v0.1.1)}, v0.1.1, Zenodo,  Zenodo,
  \dodoi{10.5281/zenodo.12157}

\bibitem[{{Fujimoto} {et~al.}(2023){Fujimoto}, {Arrabal Haro}, {Dickinson},
  {Finkelstein}, {Kartaltepe}, {Larson}, {Burgarella}, {Bagley}, {Behroozi},
  {Chworowsky}, {Hirschmann}, {Trump}, {Wilkins}, {Yung}, {Koekemoer},
  {Papovich}, {Pirzkal}, {Ferguson}, {Fontana}, {Grogin}, {Grazian}, {Kewley},
  {Kocevski}, {Lotz}, {Pentericci}, {Ravindranath}, {Somerville}, {Wilkins},
  {Amor{\'\i}n}, {Backhaus}, {Calabr{\`o}}, {Casey}, {Cooper}, {Fern{\'a}ndez},
  {Franco}, {Giavalisco}, {Hathi}, {Harish}, {Hutchison}, {Iyer}, {Jung},
  {Lucas}, \& {Zavala}}]{fujimoto2023}
{Fujimoto}, S., {Arrabal Haro}, P., {Dickinson}, M., {et~al.} 2023, \apjl, 949,
  L25, \dodoi{10.3847/2041-8213/acd2d9}

\bibitem[{{Gunawardhana} {et~al.}(2011){Gunawardhana}, {Hopkins}, {Sharp},
  {Brough}, {Taylor}, {Bland-Hawthorn}, {Maraston}, {Tuffs}, {Popescu},
  {Wijesinghe}, {Jones}, {Croom}, {Sadler}, {Wilkins}, {Driver}, {Liske},
  {Norberg}, {Baldry}, {Bamford}, {Loveday}, {Peacock}, {Robotham}, {Zucker},
  {Parker}, {Conselice}, {Cameron}, {Frenk}, {Hill}, {Kelvin}, {Kuijken},
  {Madore}, {Nichol}, {Parkinson}, {Pimbblet}, {Prescott}, {Sutherland},
  {Thomas}, \& {van Kampen}}]{gunawardhana2011}
{Gunawardhana}, M.~L.~P., {Hopkins}, A.~M., {Sharp}, R.~G., {et~al.} 2011,
  \mnras, 415, 1647, \dodoi{10.1111/j.1365-2966.2011.18800.x}

\bibitem[{{Hainline} {et~al.}(2023){Hainline}, {Johnson}, {Robertson},
  {Tacchella}, {Helton}, {Sun}, {Eisenstein}, {Simmonds}, {Topping}, {Whitler},
  {Willmer}, {Rieke}, {Suess}, {Hviding}, {Cameron}, {Alberts}, {Baker},
  {Bhatawdekar}, {Boyett}, {Bunker}, {Carniani}, {Charlot}, {Chen}, {Curti},
  {Curtis-Lake}, {D'Eugenio}, {Egami}, {Endsley}, {Hausen}, {Ji}, {Looser},
  {Lyu}, {Maiolino}, {Nelson}, {Puskas}, {Rawle}, {Sandles}, {Saxena}, {Smit},
  {Stark}, {Williams}, {Willott}, \& {Witstok}}]{Hainline2023}
{Hainline}, K.~N., {Johnson}, B.~D., {Robertson}, B., {et~al.} 2023, arXiv
  e-prints, arXiv:2306.02468, \dodoi{10.48550/arXiv.2306.02468}

\bibitem[{{Harikane} {et~al.}(2023){Harikane}, {Nakajima}, {Ouchi}, {Umeda},
  {Isobe}, {Ono}, {Xu}, \& {Zhang}}]{harikane2023}
{Harikane}, Y., {Nakajima}, K., {Ouchi}, M., {et~al.} 2023, arXiv e-prints,
  arXiv:2304.06658, \dodoi{10.48550/arXiv.2304.06658}

\bibitem[{{Helton} {et~al.}(2023){Helton}, {Sun}, {Woodrum}, {Hainline},
  {Willmer}, {Rieke}, {Rieke}, {Tacchella}, {Robertson}, {Johnson}, {Alberts},
  {Eisenstein}, {Hausen}, {Bonaventura}, {Bunker}, {Charlot}, {Curti},
  {Curtis-Lake}, {Looser}, {Maiolino}, {Willott}, {Witstok}, {Boyett}, {Chen},
  {Egami}, {Endsley}, {Hviding}, {Jaffe}, {Ji}, {Lyu}, \&
  {Sandles}}]{Helton2023}
{Helton}, J.~M., {Sun}, F., {Woodrum}, C., {et~al.} 2023, arXiv e-prints,
  arXiv:2302.10217, \dodoi{10.48550/arXiv.2302.10217}

\bibitem[{{Hopkins}(2012)}]{Hopkins2012}
{Hopkins}, P.~F. 2012, \mnras, 423, 2037,
  \dodoi{10.1111/j.1365-2966.2012.20731.x}

\bibitem[{{Jeans}(1902)}]{Jeans1902}
{Jeans}, J.~H. 1902, Philosophical Transactions of the Royal Society of London
  Series A, 199, 1, \dodoi{10.1098/rsta.1902.0012}

\bibitem[{{Jermyn} {et~al.}(2018){Jermyn}, {Steinhardt}, \&
  {Tout}}]{Jermyn2018}
{Jermyn}, A.~S., {Steinhardt}, C.~L., \& {Tout}, C.~A. 2018, \mnras, 480, 4265,
  \dodoi{10.1093/mnras/sty2123}

\bibitem[{{Johnson} {et~al.}(2021){Johnson}, {Leja}, {Conroy}, \&
  {Speagle}}]{Johnson2021}
{Johnson}, B.~D., {Leja}, J., {Conroy}, C., \& {Speagle}, J.~S. 2021, \apjs,
  254, 22, \dodoi{10.3847/1538-4365/abef67}

\bibitem[{{Kocevski} {et~al.}(2023){Kocevski}, {Onoue}, {Inayoshi}, {Trump},
  {Haro}, {Grazian}, {Dickinson}, {Finkelstein}, {Kartaltepe}, {Hirschmann},
  {Aird}, {Holwerda}, {Fujimoto}, {Juneau}, {Amor{\'\i}n}, {Backhaus},
  {Bagley}, {Barro}, {Bell}, {Bisigello}, {Calabr{\`o}}, {Cleri}, {Cooper},
  {Ding}, {Grogin}, {Ho}, {Hutchison}, {Inoue}, {Jiang}, {Jones}, {Koekemoer},
  {Li}, {Li}, {McGrath}, {Molina}, {Papovich}, {P{\'e}rez-Gonz{\'a}lez},
  {Pirzkal}, {Wilkins}, {Yang}, \& {Yung}}]{kocevski2023}
{Kocevski}, D.~D., {Onoue}, M., {Inayoshi}, K., {et~al.} 2023, \apjl, 954, L4,
  \dodoi{10.3847/2041-8213/ace5a0}

\bibitem[{{Kriek} \& {Conroy}(2013)}]{Kriek2013}
{Kriek}, M., \& {Conroy}, C. 2013, \apjl, 775, L16,
  \dodoi{10.1088/2041-8205/775/1/L16}

\bibitem[{{Labb{\'e}} {et~al.}(2023){Labb{\'e}}, {van Dokkum}, {Nelson},
  {Bezanson}, {Suess}, {Leja}, {Brammer}, {Whitaker}, {Mathews}, {Stefanon}, \&
  {Wang}}]{Labbe2022}
{Labb{\'e}}, I., {van Dokkum}, P., {Nelson}, E., {et~al.} 2023, \nat, 616, 266,
  \dodoi{10.1038/s41586-023-05786-2}

\bibitem[{{Looser} {et~al.}(2023{\natexlab{a}}){Looser}, {D'Eugenio},
  {Maiolino}, {Witstok}, {Sandles}, {Curtis-Lake}, {Chevallard}, {Tacchella},
  {Johnson}, {Baker}, {Suess}, {Carniani}, {Ferruit}, {Arribas}, {Bonaventura},
  {Bunker}, {Cameron}, {Charlot}, {Curti}, {de Graaff}, {Maseda}, {Rawle},
  {Rix}, {Rodriguez Del Pino}, {Smit}, {{\"U}bler}, {Willott}, {Alberts},
  {Egami}, {Eisenstein}, {Endsley}, {Hausen}, {Rieke}, {Robertson}, {Shivaei},
  {Williams}, {Boyett}, {Chen}, {Ji}, {Jones}, {Kumari}, {Nelson}, {Perna},
  {Saxena}, \& {Scholtz}}]{Looser2023}
{Looser}, T.~J., {D'Eugenio}, F., {Maiolino}, R., {et~al.} 2023{\natexlab{a}},
  arXiv e-prints, arXiv:2302.14155, \dodoi{10.48550/arXiv.2302.14155}

\bibitem[{{Looser} {et~al.}(2023{\natexlab{b}}){Looser}, {D'Eugenio},
  {Maiolino}, {Tacchella}, {Curti}, {Arribas}, {Baker}, {Baum}, {Bonaventura},
  {Boyett}, {Bunker}, {Carniani}, {Charlot}, {Chevallard}, {Curtis-Lake},
  {Danhaive}, {Eisenstein}, {de Graaff}, {Hainline}, {Ji}, {Johnson}, {Kumari},
  {Nelson}, {Parlanti}, {Rix}, {Robertson}, {Rodr{\'\i}guez Del Pino},
  {Sandles}, {Scholtz}, {Smit}, {Stark}, {{\"U}bler}, {Williams}, {Willott}, \&
  {Witstok}}]{Looser2023b}
---. 2023{\natexlab{b}}, arXiv e-prints, arXiv:2306.02470,
  \dodoi{10.48550/arXiv.2306.02470}

\bibitem[{{Lovell} {et~al.}(2023){Lovell}, {Harrison}, {Harikane}, {Tacchella},
  \& {Wilkins}}]{lovell2023}
{Lovell}, C.~C., {Harrison}, I., {Harikane}, Y., {Tacchella}, S., \& {Wilkins},
  S.~M. 2023, \mnras, 518, 2511, \dodoi{10.1093/mnras/stac3224}

\bibitem[{{Madau}(1995)}]{Madau1995}
{Madau}, P. 1995, \apj, 441, 18, \dodoi{10.1086/175332}

\bibitem[{{Narayanan} \& {Dav{\'e}}(2012)}]{Narayanan2012}
{Narayanan}, D., \& {Dav{\'e}}, R. 2012, \mnras, 423, 3601,
  \dodoi{10.1111/j.1365-2966.2012.21159.x}

\bibitem[{{Narayanan} {et~al.}(2023){Narayanan}, {Lower}, {Torrey}, {Brammer},
  {Cui}, {Dave}, {Iyer}, {Li}, {Lovell}, {Sales}, {Stark}, {Marinacci}, \&
  {Vogelsberger}}]{narayanan2023}
{Narayanan}, D., {Lower}, S., {Torrey}, P., {et~al.} 2023, arXiv e-prints,
  arXiv:2306.10118, \dodoi{10.48550/arXiv.2306.10118}

\bibitem[{{Paxton} {et~al.}(2011){Paxton}, {Bildsten}, {Dotter}, {Herwig},
  {Lesaffre}, \& {Timmes}}]{Paxton2011}
{Paxton}, B., {Bildsten}, L., {Dotter}, A., {et~al.} 2011, \apjs, 192, 3,
  \dodoi{10.1088/0067-0049/192/1/3}

\bibitem[{{Paxton} {et~al.}(2013){Paxton}, {Cantiello}, {Arras}, {Bildsten},
  {Brown}, {Dotter}, {Mankovich}, {Montgomery}, {Stello}, {Timmes}, \&
  {Townsend}}]{Paxton2013}
{Paxton}, B., {Cantiello}, M., {Arras}, P., {et~al.} 2013, \apjs, 208, 4,
  \dodoi{10.1088/0067-0049/208/1/4}

\bibitem[{{Paxton} {et~al.}(2015){Paxton}, {Marchant}, {Schwab}, {Bauer},
  {Bildsten}, {Cantiello}, {Dessart}, {Farmer}, {Hu}, {Langer}, {Townsend},
  {Townsley}, \& {Timmes}}]{Paxton2015}
{Paxton}, B., {Marchant}, P., {Schwab}, J., {et~al.} 2015, \apjs, 220, 15,
  \dodoi{10.1088/0067-0049/220/1/15}

\bibitem[{{Paxton} {et~al.}(2018){Paxton}, {Schwab}, {Bauer}, {Bildsten},
  {Blinnikov}, {Duffell}, {Farmer}, {Goldberg}, {Marchant}, {Sorokina},
  {Thoul}, {Townsend}, \& {Timmes}}]{Paxton2018}
{Paxton}, B., {Schwab}, J., {Bauer}, E.~B., {et~al.} 2018, \apjs, 234, 34,
  \dodoi{10.3847/1538-4365/aaa5a8}

\bibitem[{{Planck Collaboration} {et~al.}(2020){Planck Collaboration},
  {Aghanim}, {Akrami}, {Ashdown}, {Aumont}, {Baccigalupi}, {Ballardini},
  {Banday}, {Barreiro}, {Bartolo}, {Basak}, {Battye}, {Benabed}, {Bernard},
  {Bersanelli}, {Bielewicz}, {Bock}, {Bond}, {Borrill}, {Bouchet}, {Boulanger},
  {Bucher}, {Burigana}, {Butler}, {Calabrese}, {Cardoso}, {Carron},
  {Challinor}, {Chiang}, {Chluba}, {Colombo}, {Combet}, {Contreras}, {Crill},
  {Cuttaia}, {de Bernardis}, {de Zotti}, {Delabrouille}, {Delouis}, {Di
  Valentino}, {Diego}, {Dor{\'e}}, {Douspis}, {Ducout}, {Dupac}, {Dusini},
  {Efstathiou}, {Elsner}, {En{\ss}lin}, {Eriksen}, {Fantaye}, {Farhang},
  {Fergusson}, {Fernandez-Cobos}, {Finelli}, {Forastieri}, {Frailis},
  {Fraisse}, {Franceschi}, {Frolov}, {Galeotta}, {Galli}, {Ganga},
  {G{\'e}nova-Santos}, {Gerbino}, {Ghosh}, {Gonz{\'a}lez-Nuevo}, {G{\'o}rski},
  {Gratton}, {Gruppuso}, {Gudmundsson}, {Hamann}, {Handley}, {Hansen},
  {Herranz}, {Hildebrandt}, {Hivon}, {Huang}, {Jaffe}, {Jones}, {Karakci},
  {Keih{\"a}nen}, {Keskitalo}, {Kiiveri}, {Kim}, {Kisner}, {Knox},
  {Krachmalnicoff}, {Kunz}, {Kurki-Suonio}, {Lagache}, {Lamarre}, {Lasenby},
  {Lattanzi}, {Lawrence}, {Le Jeune}, {Lemos}, {Lesgourgues}, {Levrier},
  {Lewis}, {Liguori}, {Lilje}, {Lilley}, {Lindholm}, {L{\'o}pez-Caniego},
  {Lubin}, {Ma}, {Mac{\'\i}as-P{\'e}rez}, {Maggio}, {Maino}, {Mandolesi},
  {Mangilli}, {Marcos-Caballero}, {Maris}, {Martin}, {Martinelli},
  {Mart{\'\i}nez-Gonz{\'a}lez}, {Matarrese}, {Mauri}, {McEwen}, {Meinhold},
  {Melchiorri}, {Mennella}, {Migliaccio}, {Millea}, {Mitra},
  {Miville-Desch{\^e}nes}, {Molinari}, {Montier}, {Morgante}, {Moss}, {Natoli},
  {N{\o}rgaard-Nielsen}, {Pagano}, {Paoletti}, {Partridge}, {Patanchon},
  {Peiris}, {Perrotta}, {Pettorino}, {Piacentini}, {Polastri}, {Polenta},
  {Puget}, {Rachen}, {Reinecke}, {Remazeilles}, {Renzi}, {Rocha}, {Rosset},
  {Roudier}, {Rubi{\~n}o-Mart{\'\i}n}, {Ruiz-Granados}, {Salvati}, {Sandri},
  {Savelainen}, {Scott}, {Shellard}, {Sirignano}, {Sirri}, {Spencer},
  {Sunyaev}, {Suur-Uski}, {Tauber}, {Tavagnacco}, {Tenti}, {Toffolatti},
  {Tomasi}, {Trombetti}, {Valenziano}, {Valiviita}, {Van Tent}, {Vibert},
  {Vielva}, {Villa}, {Vittorio}, {Wandelt}, {Wehus}, {White}, {White},
  {Zacchei}, \& {Zonca}}]{Planck2020}
{Planck Collaboration}, {Aghanim}, N., {Akrami}, Y., {et~al.} 2020, \aap, 641,
  A6, \dodoi{10.1051/0004-6361/201833910}

\bibitem[{{Rieke} {et~al.}(2023){Rieke}, {Robertson}, {Tacchella}, {Hainline},
  {Johnson}, {Hausan}, {Ji}, {Willmer}, {Eisenstein}, {Pusk{\`a}s}, {Alberts},
  {Arribas}, {Baker}, {Baum}, {Bhatawdekar}, {Bonaventura}, {Boyett}, {Bunker},
  {Cameron}, {Carniani}, {Charlot}, {Chevallard}, {Chen}, {Curti},
  {Curtis-Lake}, {Danhaive}, {DeCoursey}, {Dressler}, {Egami}, {Endsley},
  {Helton}, {Hviding}, {Kumari}, {Looser}, {Lyu}, {Maiolino}, {Maseda},
  {Nelson}, {Rieke}, {Rix}, {Sandles}, {Saxena}, {Sharpe}, {Shivaei},
  {Skarbinski}, {Smit}, {Stark}, {Stone}, {Suess}, {Sun}, {Topping}, {Uebler},
  {Villanueva}, {Wallace}, {Williams}, {Willott}, {Whitler}, {Witstok}, \&
  {Woodrum}}]{Rieke2023}
{Rieke}, M.~J., {Robertson}, B.~E., {Tacchella}, S., {et~al.} 2023, arXiv
  e-prints, arXiv:2306.02466, \dodoi{10.48550/arXiv.2306.02466}

\bibitem[{{Robertson}(2022)}]{robertson2022}
{Robertson}, B.~E. 2022, \araa, 60, 121,
  \dodoi{10.1146/annurev-astro-120221-044656}

\bibitem[{{Robertson} {et~al.}(2023){Robertson}, {Tacchella}, {Johnson},
  {Hainline}, {Whitler}, {Eisenstein}, {Endsley}, {Rieke}, {Stark}, {Alberts},
  {Dressler}, {Egami}, {Hausen}, {Rieke}, {Shivaei}, {Williams}, {Willmer},
  {Arribas}, {Bonaventura}, {Bunker}, {Cameron}, {Carniani}, {Charlot},
  {Chevallard}, {Curti}, {Curtis-Lake}, {D'Eugenio}, {Jakobsen}, {Looser},
  {L{\"u}tzgendorf}, {Maiolino}, {Maseda}, {Rawle}, {Rix}, {Smit}, {{\"U}bler},
  {Willott}, {Witstok}, {Baum}, {Bhatawdekar}, {Boyett}, {Chen}, {de Graaff},
  {Florian}, {Helton}, {Hviding}, {Ji}, {Kumari}, {Lyu}, {Nelson}, {Sandles},
  {Saxena}, {Suess}, {Sun}, {Topping}, \& {Wallace}}]{robertson_etal23}
{Robertson}, B.~E., {Tacchella}, S., {Johnson}, B.~D., {et~al.} 2023, Nature
  Astronomy, 7, 611, \dodoi{10.1038/s41550-023-01921-1}

\bibitem[{{Salpeter}(1955)}]{Salpeter1955}
{Salpeter}, E.~E. 1955, \apj, 121, 161, \dodoi{10.1086/145971}

\bibitem[{{Sneppen} {et~al.}(2022){Sneppen}, {Steinhardt}, {Hensley}, {Jermyn},
  {Mostafa}, \& {Weaver}}]{Sneppen2022}
{Sneppen}, A., {Steinhardt}, C.~L., {Hensley}, H., {et~al.} 2022, \apj, 931,
  57, \dodoi{10.3847/1538-4357/ac695e}

\bibitem[{{Sommovigo} {et~al.}(2022){Sommovigo}, {Ferrara}, {Pallottini},
  {Dayal}, {Bouwens}, {Smit}, {da Cunha}, {De Looze}, {Bowler}, {Hodge},
  {Inami}, {Oesch}, {Endsley}, {Gonzalez}, {Schouws}, {Stark}, {Stefanon},
  {Aravena}, {Graziani}, {Riechers}, {Schneider}, {van der Werf}, {Algera},
  {Barrufet}, {Fudamoto}, {Hygate}, {Labb{\'e}}, {Li}, {Nanayakkara}, \&
  {Topping}}]{sommovigo2022}
{Sommovigo}, L., {Ferrara}, A., {Pallottini}, A., {et~al.} 2022, \mnras, 513,
  3122, \dodoi{10.1093/mnras/stac302}

\bibitem[{{Speagle}(2020)}]{dynesty:2020}
{Speagle}, J.~S. 2020, \mnras, \dodoi{10.1093/mnras/staa278}

\bibitem[{{Stark}(2016)}]{stark2016}
{Stark}, D.~P. 2016, \araa, 54, 761,
  \dodoi{10.1146/annurev-astro-081915-023417}

\bibitem[{{Steinhardt} {et~al.}(2016){Steinhardt}, {Capak}, {Masters}, \&
  {Speagle}}]{Steinhardt2016}
{Steinhardt}, C.~L., {Capak}, P., {Masters}, D., \& {Speagle}, J.~S. 2016,
  \apj, 824, 21, \dodoi{10.3847/0004-637X/824/1/21}

\bibitem[{{Steinhardt} {et~al.}(2020){Steinhardt}, {Jermyn}, \&
  {Lodman}}]{Steinhardt2020}
{Steinhardt}, C.~L., {Jermyn}, A.~S., \& {Lodman}, J. 2020, \apj, 890, 19,
  \dodoi{10.3847/1538-4357/ab66b7}

\bibitem[{{Steinhardt} {et~al.}(2023){Steinhardt}, {Kokorev}, {Rusakov},
  {Garcia}, \& {Sneppen}}]{Steinhardt2023}
{Steinhardt}, C.~L., {Kokorev}, V., {Rusakov}, V., {Garcia}, E., \& {Sneppen},
  A. 2023, \apjl, 951, L40, \dodoi{10.3847/2041-8213/acdef6}

\bibitem[{{Steinhardt} {et~al.}(2022){Steinhardt}, {Sneppen}, {Mostafa},
  {Hensley}, {Jermyn}, {Lopez}, {Weaver}, {Brammer}, {Clark}, {Davidzon},
  {Diaconu}, {Mobasher}, {Rusakov}, \& {Toft}}]{Steinhardt2022}
{Steinhardt}, C.~L., {Sneppen}, A., {Mostafa}, B., {et~al.} 2022, \apj, 931,
  58, \dodoi{10.3847/1538-4357/ac62d6}

\bibitem[{{Sun, F.} {et~al.}(2023)}]{Sun2023}
{Sun, F.}, {et~al.} 2023, {in prep.}

\bibitem[{{Tacchella} {et~al.}(2022){Tacchella}, {Finkelstein}, {Bagley},
  {Dickinson}, {Ferguson}, {Giavalisco}, {Graziani}, {Grogin}, {Hathi},
  {Hutchison}, {Jung}, {Koekemoer}, {Larson}, {Papovich}, {Pirzkal},
  {Rojas-Ruiz}, {Song}, {Schneider}, {Somerville}, {Wilkins}, \&
  {Yung}}]{tacchella2022}
{Tacchella}, S., {Finkelstein}, S.~L., {Bagley}, M., {et~al.} 2022, \apj, 927,
  170, \dodoi{10.3847/1538-4357/ac4cad}

\bibitem[{{Tacchella} {et~al.}(2023{\natexlab{a}}){Tacchella}, {Eisenstein},
  {Hainline}, {Johnson}, {Baker}, {Helton}, {Robertson}, {Suess}, {Chen},
  {Nelson}, {Pusk{\'a}s}, {Sun}, {Alberts}, {Egami}, {Hausen}, {Rieke},
  {Rieke}, {Shivaei}, {Williams}, {Willmer}, {Bunker}, {Cameron}, {Carniani},
  {Charlot}, {Curti}, {Curtis-Lake}, {Looser}, {Maiolino}, {Maseda}, {Rawle},
  {Rix}, {Smit}, {{\"U}bler}, {Willott}, {Witstok}, {Baum}, {Bhatawdekar},
  {Boyett}, {Danhaive}, {de Graaff}, {Endsley}, {Ji}, {Lyu}, {Sandles},
  {Saxena}, {Scholtz}, {Topping}, \& {Whitler}}]{Tacchella2023b}
{Tacchella}, S., {Eisenstein}, D.~J., {Hainline}, K., {et~al.}
  2023{\natexlab{a}}, \apj, 952, 74, \dodoi{10.3847/1538-4357/acdbc6}

\bibitem[{{Tacchella} {et~al.}(2023{\natexlab{b}}){Tacchella}, {Johnson},
  {Robertson}, {Carniani}, {D'Eugenio}, {Kumari}, {Maiolino}, {Nelson},
  {Suess}, {{\"U}bler}, {Williams}, {Adebusola}, {Alberts}, {Arribas},
  {Bhatawdekar}, {Bonaventura}, {Bowler}, {Bunker}, {Cameron}, {Curti},
  {Egami}, {Eisenstein}, {Frye}, {Hainline}, {Helton}, {Ji}, {Looser}, {Lyu},
  {Perna}, {Rawle}, {Rieke}, {Rieke}, {Saxena}, {Sandles}, {Shivaei},
  {Simmonds}, {Sun}, {Willmer}, {Willott}, \& {Witstok}}]{tacchella2023}
{Tacchella}, S., {Johnson}, B.~D., {Robertson}, B.~E., {et~al.}
  2023{\natexlab{b}}, \mnras, 522, 6236, \dodoi{10.1093/mnras/stad1408}

\bibitem[{{van Dokkum} {et~al.}(2023){van Dokkum}, {Brammer}, {Wang}, {Leja},
  \& {Conroy}}]{vanDokkum2023}
{van Dokkum}, P., {Brammer}, G., {Wang}, B., {Leja}, J., \& {Conroy}, C. 2023,
  arXiv e-prints, arXiv:2309.07969, \dodoi{10.48550/arXiv.2309.07969}

\bibitem[{{van Dokkum}(2008)}]{vanDokkum2008}
{van Dokkum}, P.~G. 2008, \apj, 674, 29, \dodoi{10.1086/525014}

\bibitem[{{Vazdekis} {et~al.}(2015){Vazdekis}, {Coelho}, {Cassisi},
  {Ricciardelli}, {Falc{\'o}n-Barroso}, {S{\'a}nchez-Bl{\'a}zquez}, {La
  Barbera}, {Beasley}, \& {Pietrinferni}}]{Vazdekis2015}
{Vazdekis}, A., {Coelho}, P., {Cassisi}, S., {et~al.} 2015, \mnras, 449, 1177,
  \dodoi{10.1093/mnras/stv151}

\bibitem[{{Whitler} {et~al.}(2023){Whitler}, {Stark}, {Endsley}, {Leja},
  {Charlot}, \& {Chevallard}}]{whitler2023}
{Whitler}, L., {Stark}, D.~P., {Endsley}, R., {et~al.} 2023, \mnras, 519, 5859,
  \dodoi{10.1093/mnras/stad004}

\bibitem[{{Williams} {et~al.}(2023){Williams}, {Tacchella}, {Maseda},
  {Robertson}, {Johnson}, {Willott}, {Eisenstein}, {Willmer}, {Ji}, {Hainline},
  {Helton}, {Alberts}, {Baum}, {Bhatawdekar}, {Boyett}, {Bunker}, {Carniani},
  {Charlot}, {Chevallard}, {Curtis-Lake}, {de Graaf}, {Egami}, {Franx},
  {Kumari}, {Maiolino}, {Nelson}, {Rieke}, {Sandles}, {Shivaei}, {Simmonds},
  {Smit}, {Suess}, {Sun}, {Ubler}, \& {Witstok}}]{Williams2023}
{Williams}, C.~C., {Tacchella}, S., {Maseda}, M.~V., {et~al.} 2023, arXiv
  e-prints, arXiv:2301.09780, \dodoi{10.48550/arXiv.2301.09780}

\bibitem[{{Witstok} {et~al.}(2023){Witstok}, {Shivaei}, {Smit}, {Maiolino},
  {Carniani}, {Curtis-Lake}, {Ferruit}, {Arribas}, {Bunker}, {Cameron},
  {Charlot}, {Chevallard}, {Curti}, {de Graaff}, {D'Eugenio}, {Giardino},
  {Looser}, {Rawle}, {Rodr{\'\i}guez del Pino}, {Willott}, {Alberts}, {Baker},
  {Boyett}, {Egami}, {Eisenstein}, {Endsley}, {Hainline}, {Ji}, {Johnson},
  {Kumari}, {Lyu}, {Nelson}, {Perna}, {Rieke}, {Robertson}, {Sandles},
  {Saxena}, {Scholtz}, {Sun}, {Tacchella}, {Williams}, \&
  {Willmer}}]{Witstok2023}
{Witstok}, J., {Shivaei}, I., {Smit}, R., {et~al.} 2023, \nat, 621, 267,
  \dodoi{10.1038/s41586-023-06413-w}

\end{thebibliography}
\end{document}